\documentclass[preprint,superscriptaddress,groupedaddress]{revtex4}
\pdfoutput=1

\usepackage{graphicx}
\usepackage{bmpsize}
\usepackage{bm}        % for math
\usepackage{amsmath}
\usepackage{amssymb}   % for math
\usepackage{amsopn}
\usepackage{url}
\usepackage{longtable}
\usepackage[caption=false]{subfig}
\usepackage{filecontents}
\newcommand{\SNU}{\affiliation{Seoul National University \\
Department of Physics and Astronomy \\
1 Gwanak-ro, Gwanak-gu, Seoul 08826, Korea\\
}}

\begin{document}

\title{Short-baseline Reactor Neutrino Oscillation}



\author{Seon-Hee Seo}\SNU
\email{shseo@phya.snu.ac.kr}
\date{\today}

\vspace{5.0cm}

\begin{abstract}
The successful measurements of the smallest neutrino mixing angle, $\theta_{13}$, in 2012
by the short (1$\sim$2 km) baseline reactor neutrinos experiments, Daya Bay, RENO, and Double Chooz, 
have triggered a golden age of neutrino physics. 
The three experiments have been improving the $\theta_{13}$ measurements by accumulating event statistics
and reducing systematic uncertainties. 
Now the $\theta_{13}$ measurement is the most precise one among the mixing angles in the Pontecorvo-Maki-Nakagawa-Sakata matrix.
The most updated $\theta_{13}$ and $\Delta m^{2}_{ee}$ measurements from these experiments are reported here
as well as the 5 MeV excess, absolute reactor neutrino flux and sterile neutrino search.
The best final precision on the sin$^{2}2\theta_{13}$ ($|\Delta m^2_{ee}|$) measurement is expected to be $\sim$3\% ($\sim$3\%). 
A combined analysis from the three experiments will reduce the uncertainty and the relevant activity has started recently.
\end{abstract}

\maketitle

% Keep total 6 pages (excluding the front page)
%=====================================================
\section{Introduction}

Reactor neutrinos have played important role in neutrino physics starting from the discovery of the neutrinos in 1954 by Reines and Cowan group~\cite{Reines}
to definitive measurement of $\theta_{13}$ in 2012 by the short-baseline (1$\sim$2 km) reactor neutrino experiments, Daya Bay~\cite{DYB_2012} and RENO~\cite{RENO_2012}. 
A reactor is a copious source of electron anti-neutrinos ($\overline{\nu}_{e}$) producing $\sim$2.2$\times$10$^{20} \overline{\nu}_{e}$ per GW$_{th}$. 
The total thermal powers from the reactors of these experiments are 17.4 GW$_{th}$ (Daya Bay), 16 GW$_{th}$ (RENO), and 8.5 GW$_{th}$ (Double Chooz). 
The $\overline{\nu}_{e}$ survival probability~\cite{Petcov} is written as
\begin{eqnarray}
 P(\overline{\nu}_{e} \rightarrow \overline{\nu}_{e}) & =  & 1 - \sin^2 2 \theta_{13} ( \cos^2 \theta_{12} \sin^2 \Delta_{31} + \sin^2 \theta_{12} \sin^2 \Delta_{32} )
  \nonumber 
  - \cos^4 \theta_{13} \sin^2 2\theta_{12} \sin^2 \Delta_{21}
  \nonumber       \\
 &  &  \hspace*{-1.3cm} ~ \approx  1 - \sin^2 2 \theta_{13} \sin^2 \Delta_{ee}  - \cos^4 \theta_{13} \sin^2 2\theta_{12} \sin^2 \Delta_{21},
~\label{e:Pee}
\end{eqnarray}
where $\Delta_{ij} \equiv 1.267 \Delta m_{ij}^2 L/E_{\nu}$, $E_{\nu}$ is the $\overline{\nu}_e$ energy in MeV, $L$ is the distance between the reactor and detector in meters,
and $\Delta m_{ee}^2$ is the effective neutrino mass squared difference in eV$^{2}$ and defined as
$\Delta m_{ee}^2 \equiv \cos^2 \theta_{12}\Delta m_{31}^2 + \sin^2 \theta_{12} \Delta m_{32}^2$~\cite{Parke}. 
By measuring deficit of $\overline{\nu}_{e}$ at a short base-line the $\theta_{13}$ can be measured. 

Daya Bay, RENO and Double Chooz experiments used basically the same experimental technique to measure $\theta_{13}$.
They have used liquid scintillator as neutrino target and detected positron and neutron from the Inverse Beta Decay (IBD) process: $\overline{\nu}_e + p \rightarrow e^{+} + n$.
The positron gives a prompt signal and the neutron gives a delayed signal when captured by neutron-philic elements like Gadolinium or Hydrogen
with delay time of $\sim$30 ($\sim$200) $\mu$s for neutron capture on Gd (H). 

The three experiments have adopted cylindrical shape detectors consisting of four different layers of concentric cylinder vessels.
Each region is filled with different liquids and is named as target, gamma-catcher, buffer and veto from the inner-most to outer-most order.
Target is filled with Gd doped (0.1\%) liquid scintillator in an acrylic vessel, gamma-catcher is filled with undoped liquid scintillator in an acrylic vessel, 
buffer is filled with mineral oil in a iron vessel where photo-multiplier tubes (PMTs) are attached, and veto is filled with purified water in a concrete cavity. 
Table~\ref{t:exp} summarizes the detector components for each experiment. 

\begin{table}[b]
\begin{center}
\caption{\label{t:exp}
Inner detector components and reactor total thermal power for Daya Bay, RENO and Double Chooz.
}
\begin{tabular*}{1.0\textwidth}{@{\extracolsep{\fill}} l l l c c c }
\hline
\hline
Region & Vessel & Liquids & Daya Bay & RENO & Double Chooz\\
\hline
Target & acrylic & liquid scint. (0.1\% Gd) & 20 ton ($\times$4) & 16 ton & 10 ton \\
%\hline
Gamma-catcher & acrylic & un-doped liquid scint. & 20 ton ($\times$4) & 30 ton & 20 ton \\
%\hline
Buffer & iron & mineral oil & 37 ton & 65 ton & 100 ton \\ 
\hline
\hline
Thermal power & & & 17.4 GW$_{th}$ & 16 GW$_{th}$ & 8.5 GW$_{th}$ \\
\hline
\hline
\end{tabular*}
\end{center}
\end{table}

Since it has been well known that there is a big ($\sim$6\%) uncertainty in reactor neutrino flux~\cite{flux_uncert} 
it is required to build two identical detectors by locating one at near and the other at far sites to be able to measure the smallest neutrino mixing angle $\theta_{13}$. 
The detectors are located underneath a hill to reduce spallation background. 
Table~\ref{t:overburden} shows overburden of each detector from the three experiments. 
By performing far to near ratio measurement the systematic uncertainty on the reactor neutrino flux is reduced. 
All three experiments have built two identical detectors and RENO is the first reactor neutrino experiment which started taking data using both detectors in 2011. 
\begin{table}
\begin{center}
\caption{\label{t:overburden}
Overburden of each detector from Daya Bay, RENO and Double Chooz.
The unit m.w.e represents meter water equivalent. 
}
\begin{tabular*}{1.0\textwidth}{@{\extracolsep{\fill}} c c c c}
\hline
\hline
Experiments & Daya Bay & RENO & Double Chooz\\
\hline
Near & 250, 265 m.w.e  & 120 m.w.e & 120 m.w.e \\
Far &  860 m.w.e & 450  m.w.e & 300 m.w.e \\
\hline
\hline
\end{tabular*}
\end{center}
\end{table}

%=====================================================
\section{The $\theta_{13}$ and $|\Delta m^2_{ee}|$ measurements}

\begin{figure}[b]
\begin{center}
\includegraphics[width=0.49\textwidth]{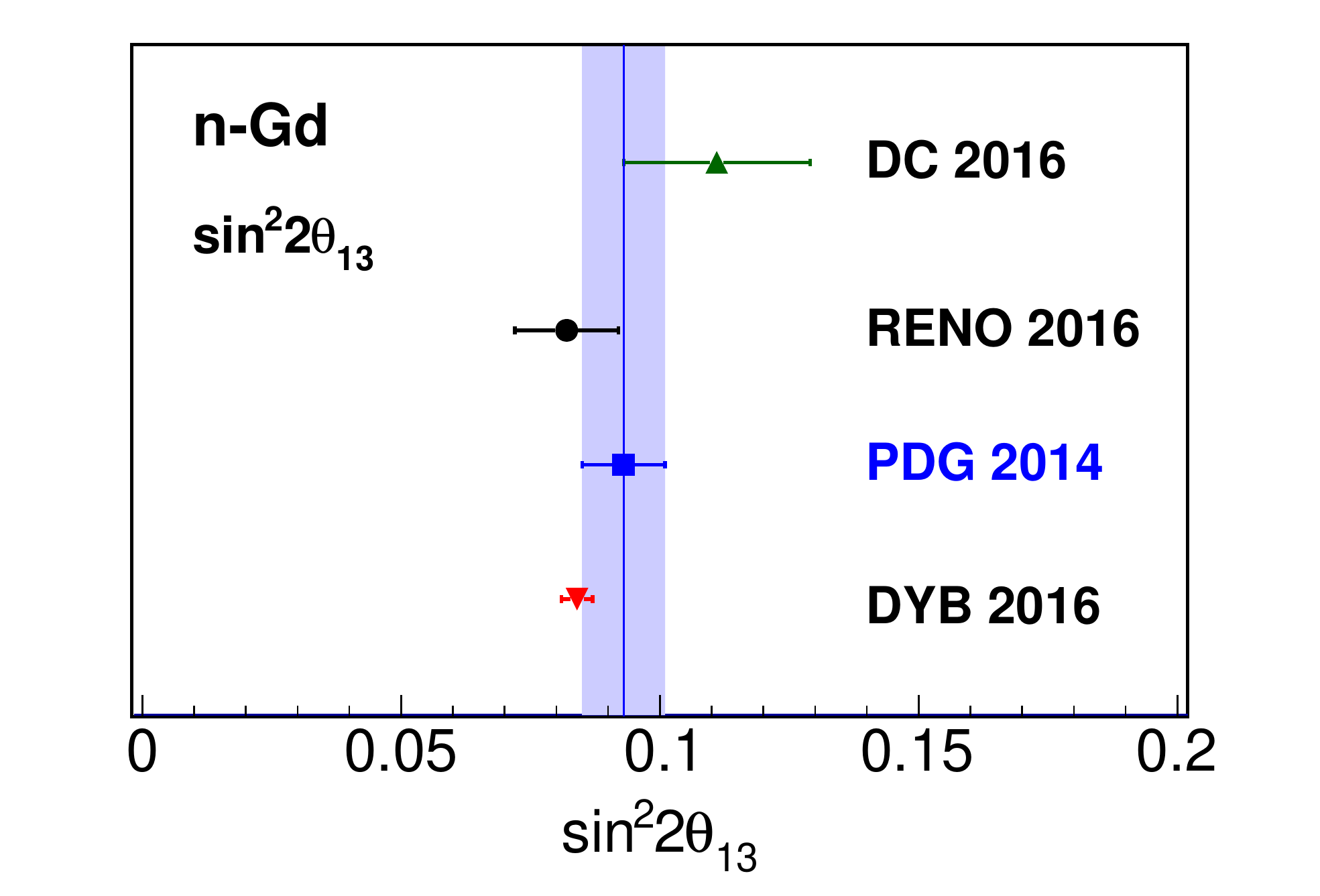}
\includegraphics[width=0.49\textwidth]{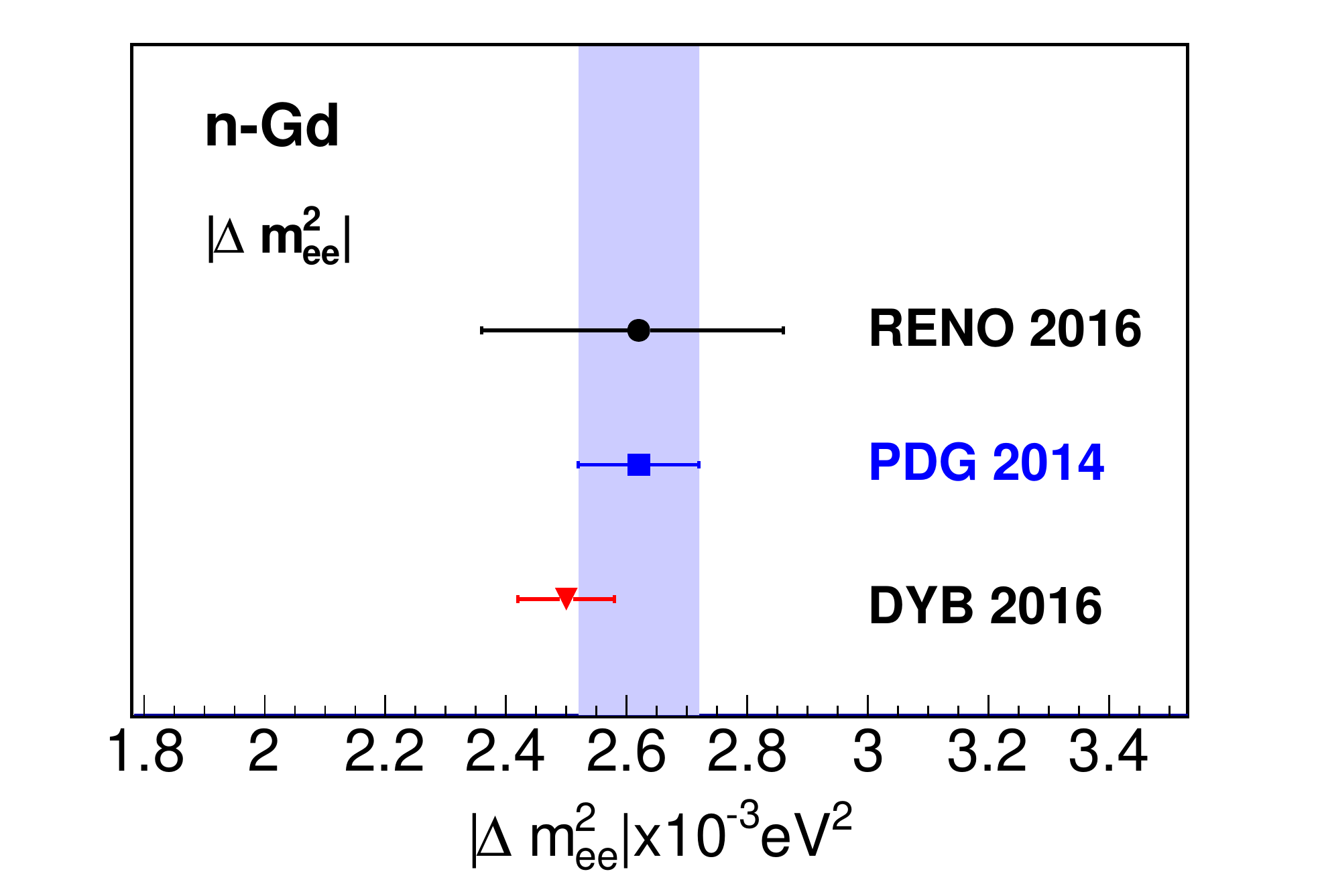}
\end{center}
\caption{
(Colors online)
The latest $\theta_{13}$ and $\Delta m^{2}_{ee}$ measurements from IBD n-Gd analyses by 
Daya Bay~\cite{DYB_1230d_2016}, RENO~\cite{RENO_nGd_2016} and Double Chooz~\cite{DC_nGd_2016}.
}
\label{f:comp}
\end{figure}

Since after the first discovery measurement of $\theta_{13}$ in 2012, more precise measurements have been done
by increasing event statistics and reducing systematic uncertainties by Daya Bay~\cite{DYB_1230d_2016,DYB_nH_2016}, RENO~\cite{RENO_nGd_2016,RENO_Neutrino2016}, 
and Double Chooz~\cite{DC_nGd_2016,DC_nH_2016}. There are two types of independent measurements depending on neutron capture on Gd (n-Gd)
or on H (n-H). Table~\ref{t:t13} and Fig.~\ref{f:comp} summarize the latest results from the three experiments
compared with the ones reported in PDG 2014~\cite{PDG2014}. 

\begin{table}
\begin{center}
\caption{\label{t:t13} 
The most updated sin$^{2}2\theta_{13}$ and $|\Delta m^2_{ee}|$ measurements from IBD n-Gd and n-H analyses by 
Daya Bay~\cite{DYB_1230d_2016,DYB_nH_2016}, RENO~\cite{RENO_nGd_2016,RENO_Neutrino2016} and Double Chooz~\cite{DC_nGd_2016,DC_nH_2016}. 
}
\begin{tabular*}{1.0\textwidth}{@{\extracolsep{\fill}} c c c c c}
\hline
\hline
Type & Experiments & Daya Bay & RENO & Double Chooz\\
\hline
n-Gd & sin$^{2}2\theta_{13}$ & 0.084$\pm$0.003 & 0.082$\pm$0.010 & 0.111$\pm$0.018 \\
\hline
n-Gd & $|\Delta m^2_{ee}|$ & [2.50$\pm$0.08]$\times$10$^{-3}$eV$^{2}$ & [2.62$^{+0.24}_{-0.26}$]$\times$10$^{-3}$eV$^{2}$ & $--$ \\
\hline
n-H & sin$^{2}2\theta_{13}$ & 0.071$\pm$0.011 & 0.086$\pm$0.019 & 0.095$^{+0.038}_{-0.039}$ \\
\hline
\hline
\end{tabular*}
\end{center}
\end{table}

%=====================================================
\section{The 5 MeV excess}

The first indication of the 5 MeV excess was raised by RENO in 2012~\cite{5MeV_RENO_2012}
and then quantitatively shown for the first time by RENO in 2014~\cite{5MeV_RENO_2014}
where RENO claimed that the 5 MeV excess is from the reactor neutrinos not predicted by Mueller and Huber model~\cite{Mueller,Huber}.
Double Chooz also showed the evidence of the 5 MeV excess in 2014~\cite{5MeV_DC_2014}
and later Daya Bay also showed the 5 MeV excess~\cite{5MeV_DB_2014}. 

\begin{figure}
\begin{center}
\includegraphics[width=0.50\textwidth]{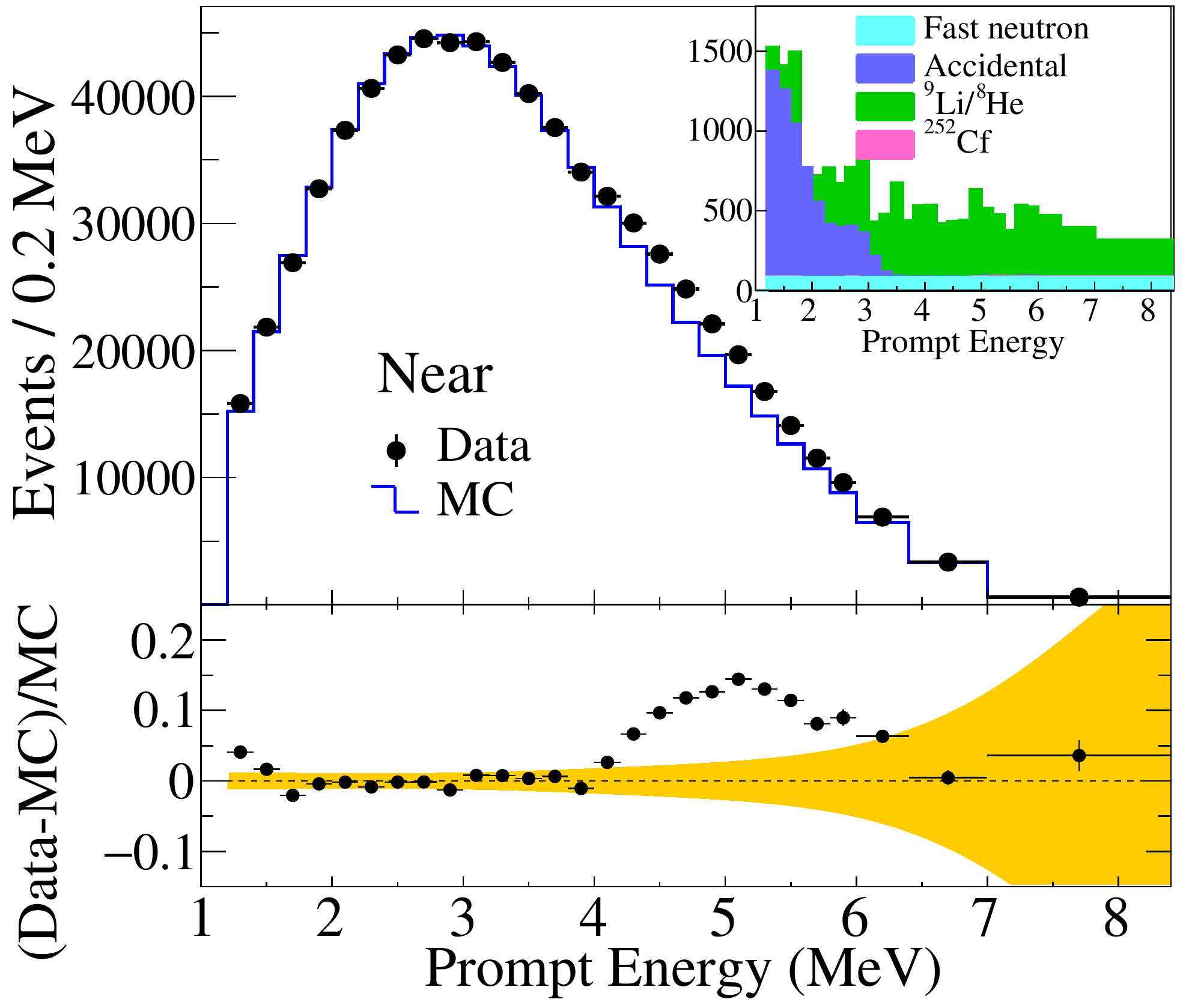}
\includegraphics[width=0.46\textwidth]{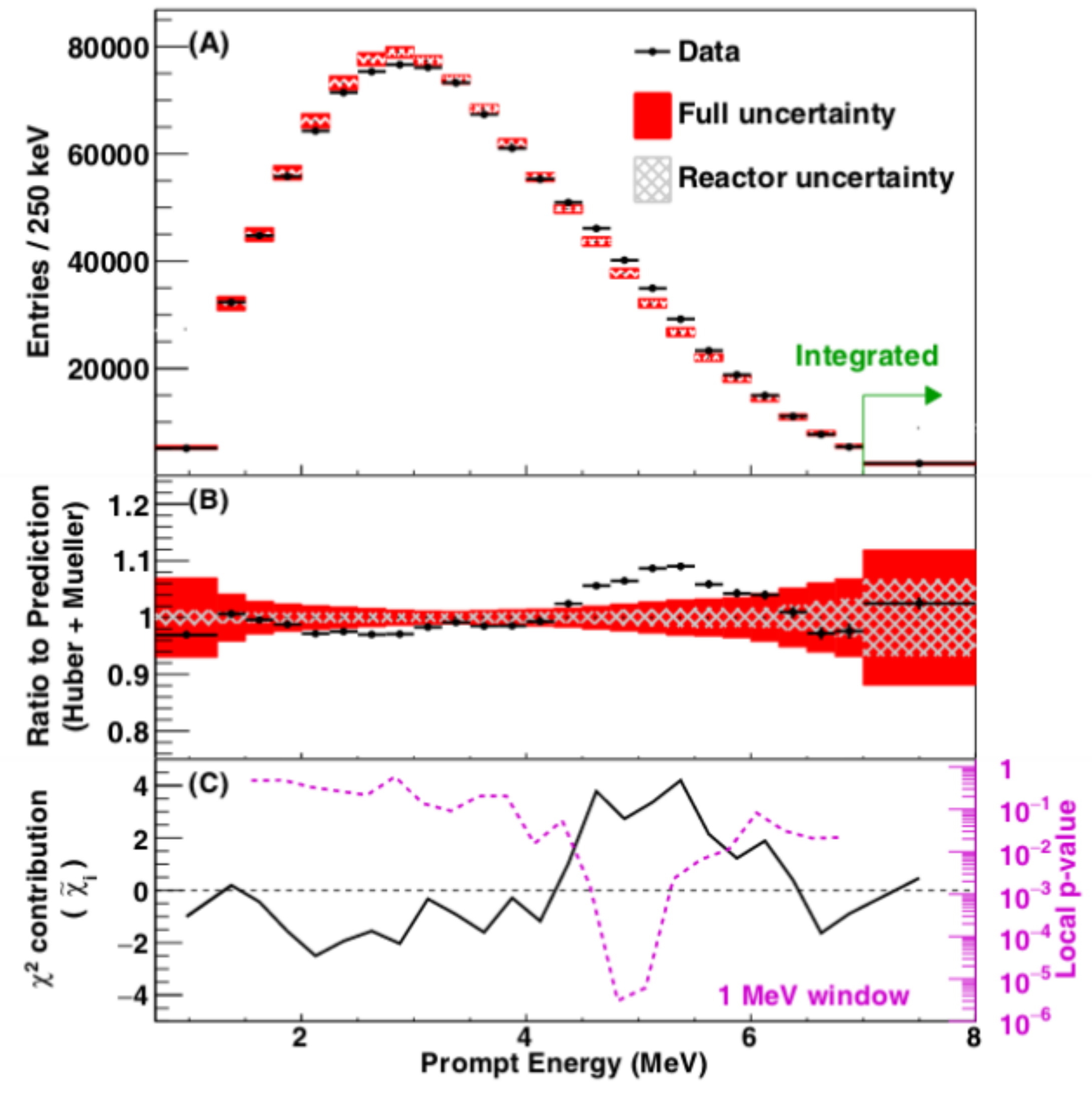}
\end{center}
\caption{
(Colors online)
The 5 MeV excesses observed by RENO~\cite{RENO_Neutrino2016} (left panel) and Daya Bay~\cite{DYB_flux_2017} (right panel). 
}
\label{f:5MeV}
\end{figure}

\begin{figure}
\begin{center}
\includegraphics[width=0.6\textwidth]{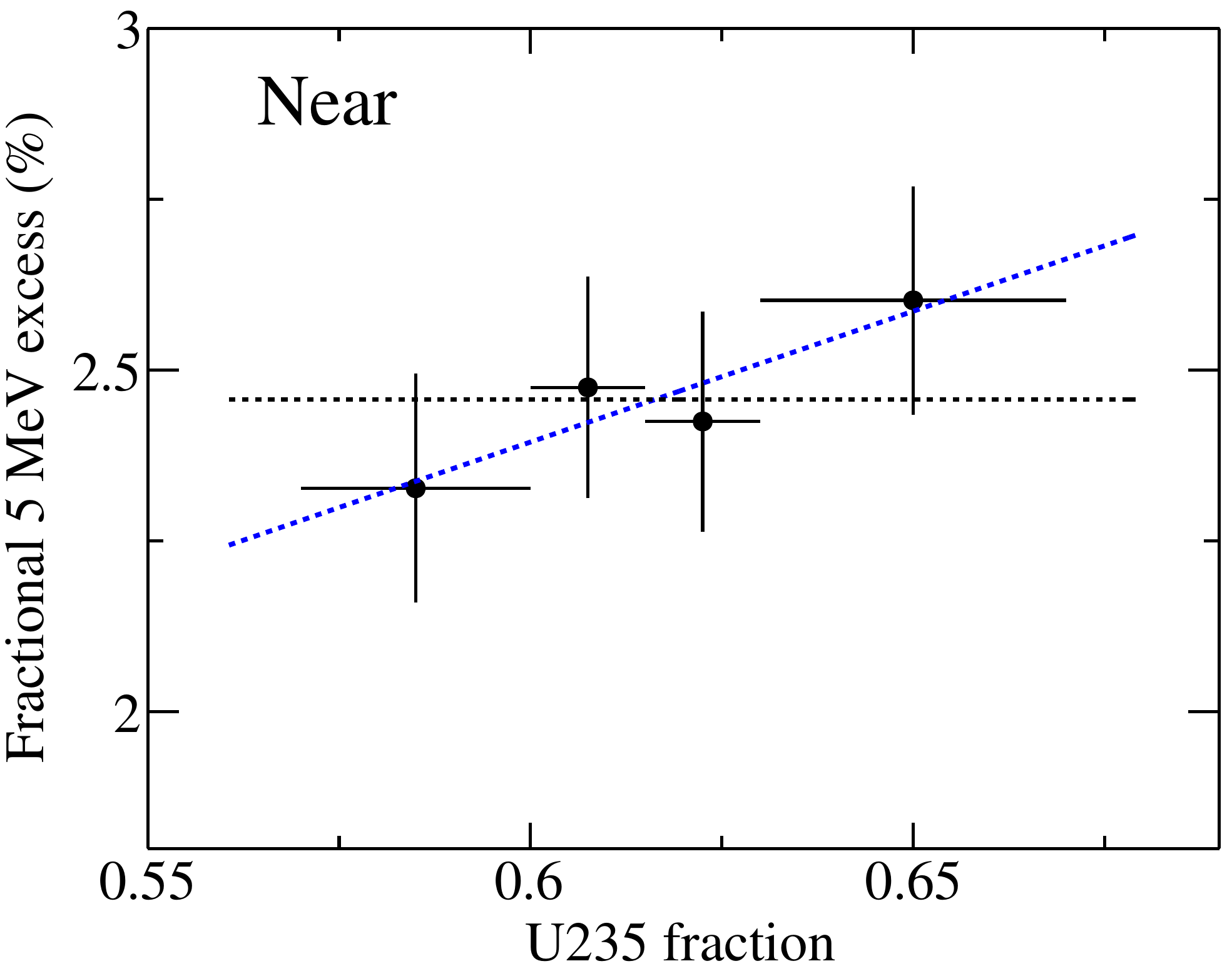}
\end{center}
\caption{
(Colors online)
Correlation between the 5 MeV excess and $^{235}$U fraction by RENO near detector~\cite{RENO_Neutrino2016}. 
The current uncertainty needs to be reduced to draw any conclusion. 
}
\label{f:corr}
\end{figure}

The most updated 5 MeV excess results are shown in Fig.~\ref{f:5MeV}
and they are 9$\sigma$ significance at RENO and 4.4$\sigma$ (3.0$\sigma$) local (global) significance at Daya Bay. 
RENO has tired to identify the correlation between the 5 MeV excess and $^{235}$U fraction as shown in Fig.~\ref{f:corr}. 
The black (blue) dotted line represents flat (first order polynomial) fitting  
but the uncertainty is currently too big to make any conclusion. 
More details on the 5 MeV excess are discussed in~\cite{dwyer,hayes,hayes2,sonzogni,huber2016}.

%=====================================================
\section{Absolute Reactor Neutrino Flux}

\begin{figure}
\begin{center}
\includegraphics[width=0.9\textwidth]{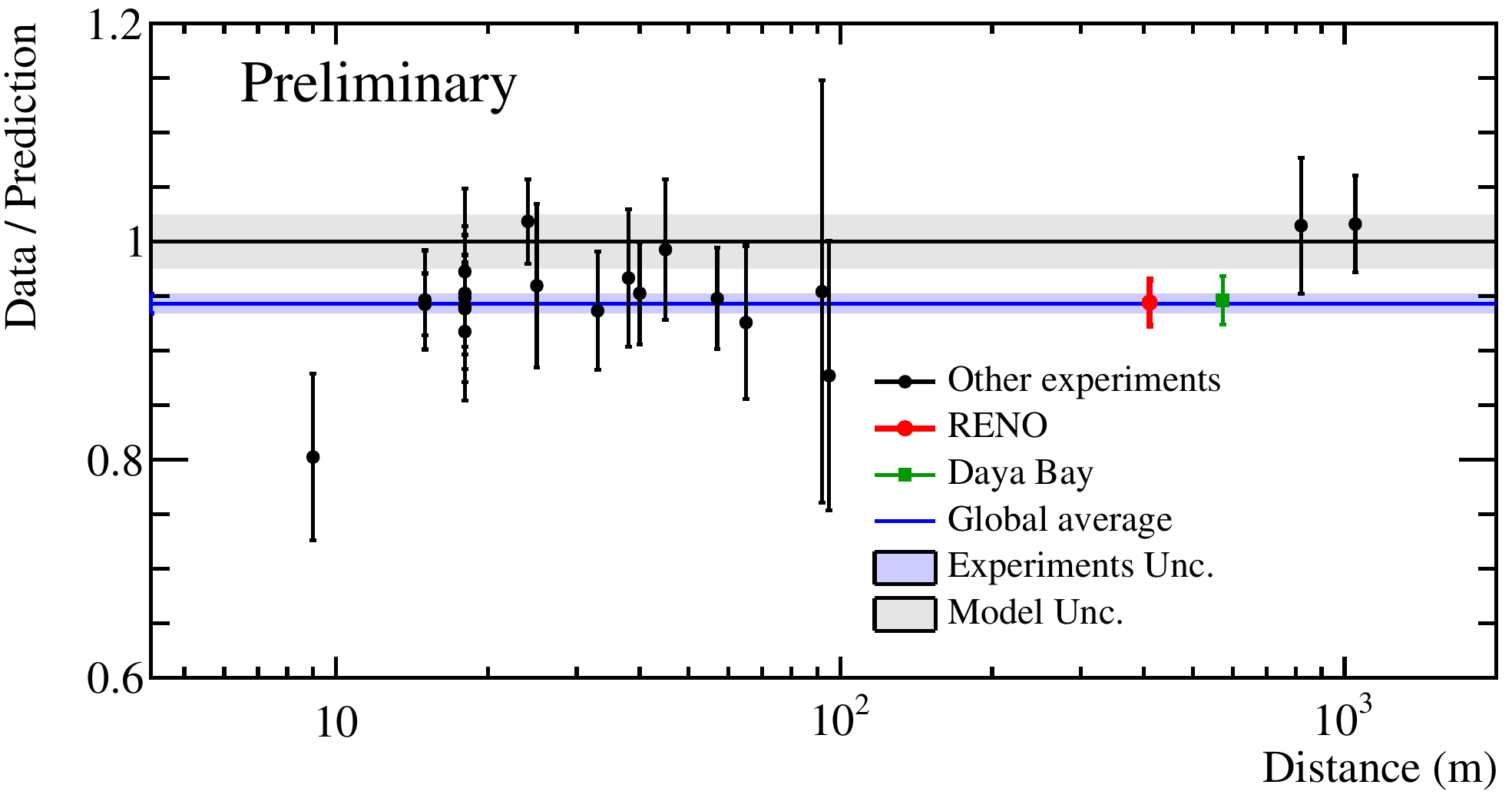}
\end{center}
\caption{
(Colors online)
Absolute neutrino fluxes. The measured points are systematically below the prediction~\cite{Mueller,Huber}. 
RENO and Daya Bay results agree well with previous measurements by very short-baseline reactor neutrino experiments. 
}
\label{f:absFlux}
\end{figure}

According to the very short-baseline ($<$ 100 m) reactor experiments the absolute reactor neutrino flux are measured to be less 
than what is expected from Mueller and Huber model~\cite{Mueller,Huber}.
The deficit could be interpreted as a sterile neutrino oscillation~\cite{flux_uncert}. 
Daya Bay~\cite{DYB_flux_2017} and RENO~\cite{RENO_Neutrino2016} also measured the absolute reactor neutrino flux and they independently obtain about 3 $\sigma$ deficit
from the Mueller and Huber model. Their measurements are 0.946$\pm$0.020 (Daya Bay) and 0.946$\pm$0.021 (RENO). 
Figure~\ref{f:absFlux} shows the absolute reactor neutrino flux measurements from Daya Bay and RENO as well as the very short-baseline reactor experiments. 

%=====================================================
\section{Sterile Neutrino Search}

\begin{figure}
\begin{center}
\includegraphics[width=0.55\textwidth]{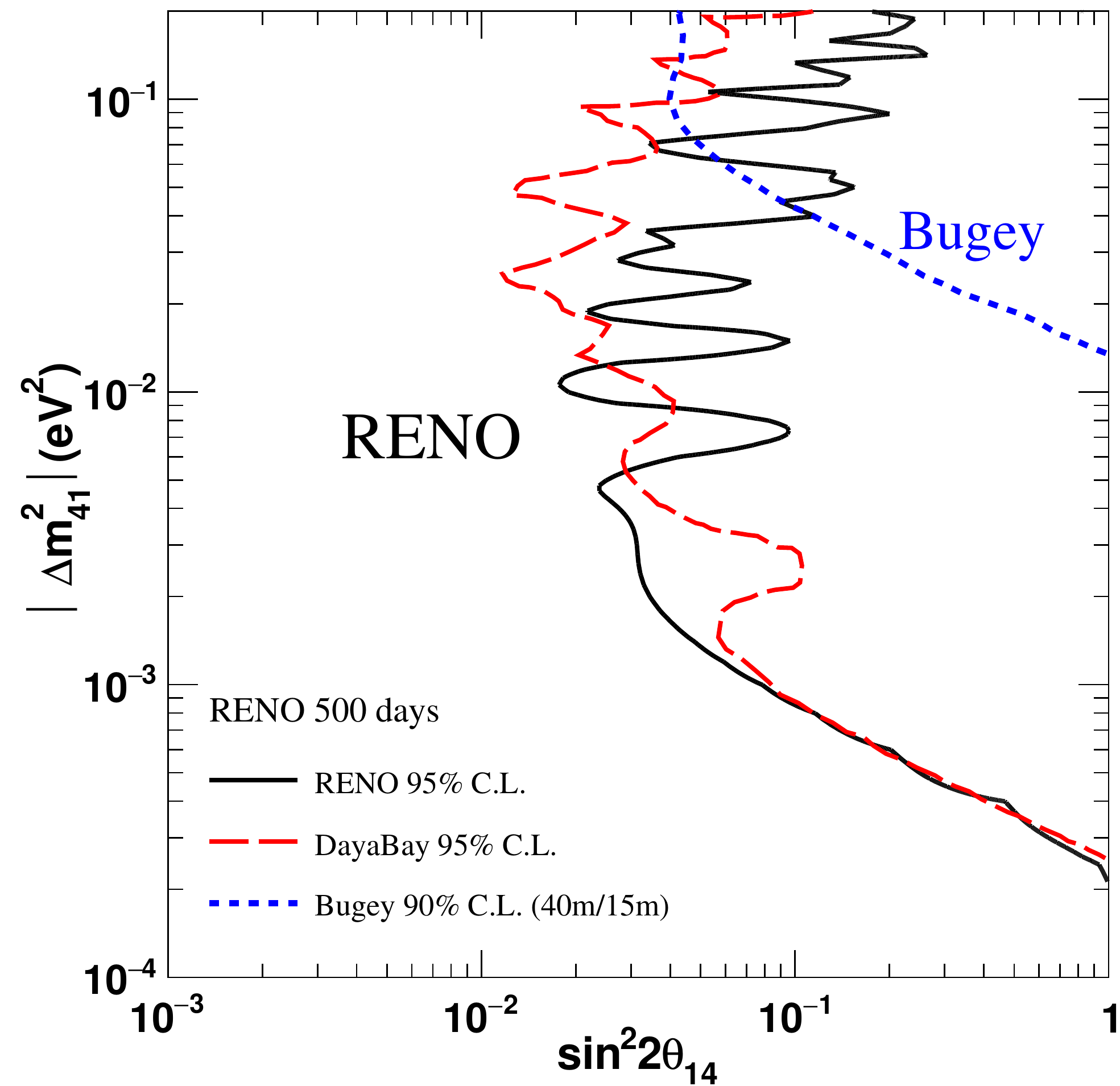}
\end{center}
\caption{
(Colors online)
Sterile neutrino search exclusion (90\% CL) regions by Daya Bay (red dashed line)~\cite{DB_sterile_2016} and RENO (black solid line)~\cite{RENO_Neutrino2016}.
}
\label{f:sterile}
\end{figure}

Four (3+1) neutrino oscillation scheme can be applied to the same data set used in the $\theta_{13}$ analysis based on three neutrino oscillation scheme. 
According to Daya Bay~\cite{DB_sterile_2016} and RENO~\cite{RENO_Neutrino2016} using the four neutrino oscillation scheme
no evidence of sterile neutrino is found, and their exclusion regions are determined as shown in Fig.~\ref{f:sterile}. 

%=====================================================
\section{Summary and Prospects}

The short-baseline reactor neutrino experiments have been very successful at measuring the smallest neutrino mixing angle 
$\theta_{13}$. 
The latest measured values on $\theta_{13}$ and $|\Delta m^2_{ee}|$ are summarized in Table~\ref{t:t13}. 
Precise measurement on $\theta_{13}$ is important since it affects the measurements of other oscillation parameters 
and can be achieved by increasing event statistics and reducing systematic uncertainties. 
Table~\ref{t:t13_future} shows the future prospects on the sin$^{2}2\theta_{13}$ and $|\Delta m^2_{ee}|$ measurements by Daya Bay, RENO and Double Chooz.
Combining the results from the three experiments might be possible so that the uncertainty on $\theta_{13}$ measurement can be further reduced.  
To discuss such a possibility the three experiments have had the first workshop in Seoul in 2016. 
This $\theta_{13}$ combined analysis workshop is expected to be continued in the near future. 

\begin{table}[h]
\begin{center}
\caption{\label{t:t13_future}
Future prospects on sin$^{2}2\theta_{13}$ and $|\Delta m^2_{ee}|$ measurements by Daya Bay, RENO and Double Chooz. 
}
\begin{tabular*}{1.0\textwidth}{@{\extracolsep{\fill}} c c c c}
\hline
\hline
Experiments & Daya Bay & RENO & Double Chooz\\
\hline
Total data &  6 years & 5 years  & 3 years for two detectors  \\
%\hline
sin$^{2}2\theta_{13}$  & $\sim$3\% &  $\sim$5\% & $\sim$10\% \\
%\hline
$|\Delta m^2_{ee}|$  &  $\sim$3\% & $\sim$5\% & $--$ \\
\hline
\hline
\end{tabular*}
\end{center}
\end{table}

The observation of the 5 MeV excess was a serendipity that enabled many active researches on understanding reactor neutrino flux model.
Currently it is not clear there is any correlation between the 5 MeV excess and $^{235}$U fraction according to RENO.
Both Daya Bay and RENO showed deficit of absolute reactor neutrino flux in 3 $\sigma$ level consistent with the previous
very short-baseline reactor neutrino measurements. Using 3+1 neutrino oscillation scheme both Daya Bay and RENO have not
found any evidence of sterile neutrinos and set 95\% CL exclusion regions in sin$^{2}2\theta_{14}$ and $\Delta m^2_{14}$ space.

%==========================================================
%

\end{document}